\documentclass[manuscript,screen]{acmart}
\AtBeginDocument{%
  \providecommand\BibTeX{{%
    \normalfont B\kern-0.5em{\scshape i\kern-0.25em b}\kern-0.8em\TeX}}}

\setcopyright{acmcopyright}
\copyrightyear{2023}
\acmYear{2023}
\acmDOI{XXXXXXX.XXXXXXX}

\acmConference[CHI '23]{ACM Conference on Human Factors in Computing Systems}{April 23--28, 2023}{Hamburg, Germany}
\acmBooktitle{CHI '23: ACM Conference on Human Factors in Computing Systems,
April 23--28, 2023, Hamburg, Germany}

\begin{document}

\title{Understanding and Mitigating Mental Health Misinformation on Video Sharing Platforms}

\author{Viet Cuong Nguyen}
\affiliation{%
  \institution{Georgia Institute of Technology}
  \streetaddress{}
  \city{}
  \country{United States of America}}
\email{johnny.nguyen@gatech.edu}
\author{Michael Birnbaum}
\affiliation{%
  \institution{Northwell Health}
  \streetaddress{}
  \city{}
  \country{United States of America}}
\email{michaelbirnbaum@northwell.edu}
\author{Munmun De Choudhury}
\affiliation{%
  \institution{Georgia Institute of Technology}
  \streetaddress{}
  \city{}
  \country{United States of America}}
\email{mchoudhu@cc.gatech.edu}

\renewcommand{\shortauthors}{Nguyen et al.}



\keywords{video-sharing platforms, social media platforms, misinformation, mental health}


\maketitle

\begin{figure}[]
    \centering
    \includegraphics[scale = 0.15]{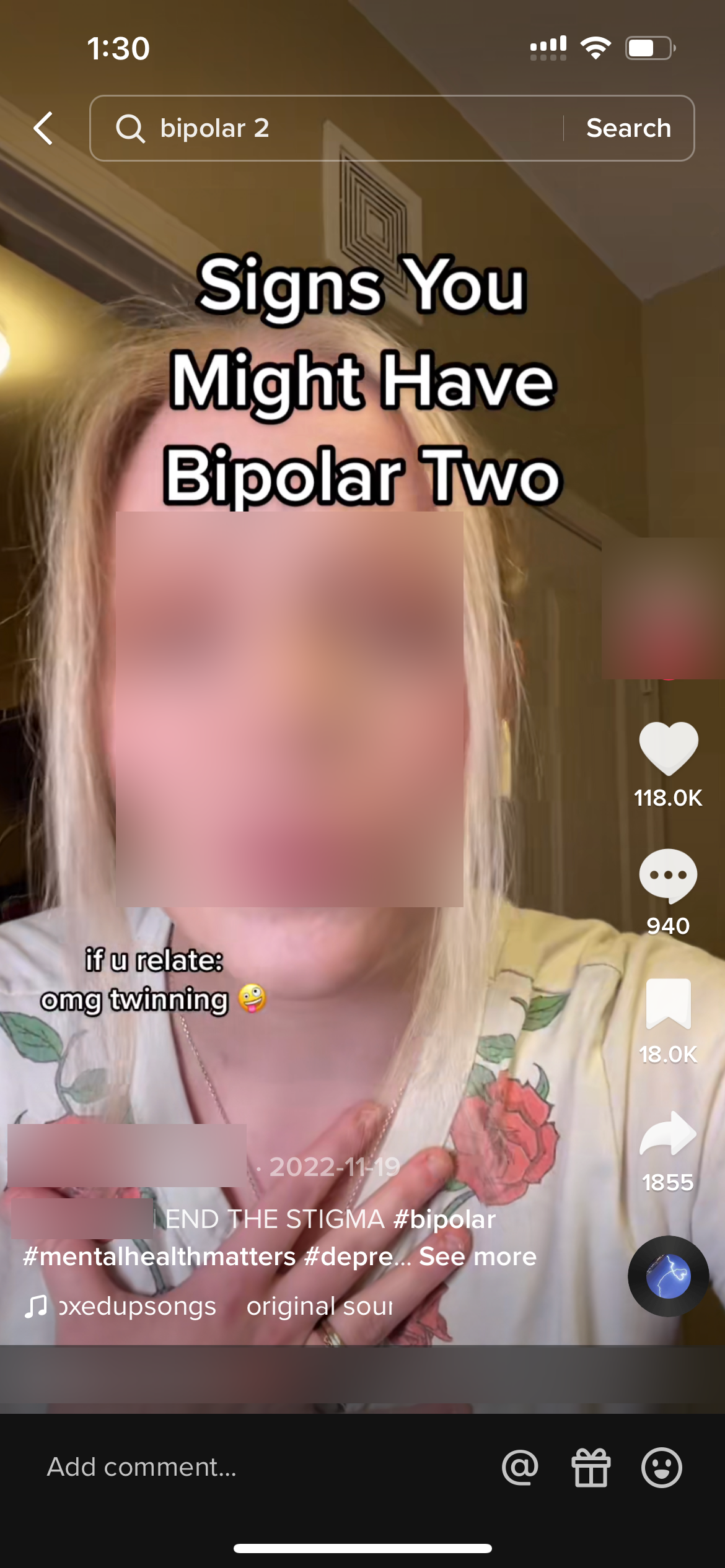}
    \caption{Example of mental health misinformation on a video-sharing platform. (potentially identifying information has been blurred out for privacy reasons)}
    \label{fig:misinfo}
\end{figure}

\section{Introduction}
Despite the ever-strong demand for mental health care globally, access to traditional mental health services remains severely limited expensive, and stifled by stigma and systemic barriers \cite{wang2005failure, wainberg2017challenges, andrade2014barriers}. Thus, over the last few years, young people are increasingly turning to content on video-sharing platforms (VSPs) like TikTok and YouTube to help them navigate their mental health journey \cite{hunter2022tiktokA, hunter2022tiktokB}. Such content is not only readily accessible and free of charge, but they contain easily digestible information through audio and visual affordances within these platforms \cite{xia2022millions, zhu2020health}. If done right, content on video-sharing platforms can be a significant asset to the growing field of digital mental health, as it can provide non-judgmental and democratized access to mental health help and advice to all, within the privacy of their realms \cite{tal2017digital, torous2020digital}. However, navigating towards trustworthy information relating to mental health on these platforms is challenging, given the uncontrollable and unregulated growth of dedicated mental health content and content creators catering to a wide array of mental health conditions on these platforms. One reason for this is the relatively low barrier of entry in creating mental health content on video-sharing platforms compared to other online-based resources (such as blog posts). Consequently, many reports have emerged that mental-health-related videos containing misinformation (referred to henceforth as \textit{mental health misinformation)} are rampant on these platforms \cite{plushcare2022tiktok, hunter2022tiktokA}. Here, we define mental health misinformation as false or misleading information about mental health and illness, including diagnosis and treatment of these challenges, irrespective of the intention of those spreading such information. An example of mental health misinformation regarding bipolar found on video-sharing platforms is shown in Figure 1. In this publicly available TikTok video where the screenshot from Figure 1 is taken, the presenter presents anecdotal symptoms which they suggest are indicative of type 2 bipolar. This video contains several key markers indicative of mental health misinformation:
\begin{itemize}
    \item They do not have relevant medical qualifications to back these statements, nor do they disclose their lack of qualifications anywhere within the video or its description
    \item The symptoms they shared for type 2 bipolar are purely anecdotal and not backed by any official diagnostic criteria for the condition (e.g. DSM-5)
    \item They encourage viewers to self-diagnose themselves with type 2 bipolar by prefacing the video with "Signs You Might Have Bipolar Two" 
\end{itemize}

There has been extensive work has been done on understanding and mitigating misinformation content on text- and image-based social media platforms such as Facebook, Twitter, and Instagram for a variety of topics \cite{grinberg2019fake, suarez2021prevalence, geeng2020fake, pierri2022propaganda, mena2020misinformation, shang2021multimodal}. However, while the virality and popularity of content on video-based social media platforms are significant, few works have focused on understanding how and mitigating the spread of mental health misinformation on video-sharing platforms. The widespread dissemination of mental health misinformation can have serious consequences. Broadly speaking, these include stigma and discrimination towards those with mental illness, the perpetuation of harmful beliefs and practices related to mental health care, misdiagnosis, and delay or avoidance of effective treatment \cite{swire2019public, devendorf2020depression}. For example, the spread of mental health misinformation on TikTok has led to the recent rise in the self-diagnosis of severe mental health disorders such as bipolar disorder, ADHD, etc. \cite{avella2023tiktok}. Thus, we argue that understanding and mitigating mental health misinformation on video-sharing platforms is crucial for reducing such dangerous self-diagnosis. In addition, doing so will ideally increase trust and credibility in video-sharing platforms as a reliable and accurate resource for digital mental health among young people. However, there are many challenges and open questions toward achieving such an ideal.    

\section{Related Works}
\subsection{Misinformation on Video-sharing Platforms}
Given the popularity of video-sharing platforms in recent years, there is an emerging thread of research focusing on misinformation hosted within these platforms. Works lying on this thread have focused on measuring the prevalence of topical misinformation content \cite{li2020youtube, donzelli2018misinformation}, detecting filter bubbles that lead viewers down a rabbit hole of misinformation content \cite{tomlein2021audit, hussein2020measuring}, and detecting misinformation videos through automated means \cite{xie2022interpretable, serrano2020nlp}.  
\subsection{Mental Health (Mis)information on Video-sharing Platforms}

Compared to other topics, academic research on mental health information and misinformation on video-sharing platforms are sparser, and to the best of our knowledge, limited to content analysis of the relevant videos \cite{yeung2022tiktok, basch2022deconstructing}. For instance, Yeung et al. [2022] examined the top 100 popular videos on TikTok regarding ADHD \cite{yeung2022tiktok}. They found that more than half of the videos were misleading, whereas only around 20 percent of the videos were labeled as 'useful'

\section{Open Questions}
Synthesizing the related works presented above, we now present some open questions we think are important to the understanding and mitigation of mental health misinformation on video-sharing platforms
\begin{itemize}
    \item How does “bad” mental health content differ from “good” mental health content in terms of virality? framing? user responses? Relatedly, how do we define what constitutes “good” and “bad” mental health information, given the rapidly evolving science in this field?
    \item How do platform affordances impact the spread of mental health misinformation on them?
    \item How can we design effective interventions against mental health misinformation on video-sharing platforms? On this note, what are the strengths and limitations of conventional content moderation approaches when applied to this sensitive domain? What moderation strategies would work and which ones wouldn’t?
    \item How can we build effective machine learning models for detecting videos containing mental health misinformation on video-sharing platforms?
    \item How do recommendation algorithms embedded in video-sharing platforms (e.g. TikTok's For You Page, Youtube's Autoplay) affect people's search for mental health content? Do watching mental health misinformation on these platforms lead young people down a rabbit hole?
    \item How do we empower psychiatrists in fact-checking mental health misinformation on video-sharing platforms
    \item How do we ensure online platforms continue to remain safe spaces for seeking and providing mental health help, with undertaking efforts that curb or reduce the impact of misinformation on these issues?  
    \item Could efforts to mitigate the harms of mental health misinformation have chilling effects on those who intend to help others with mental health struggles online? 
\end{itemize}

We believe research projects that properly address these open questions will play a crucial role towards understanding and mitigating mental health misinformation on video-sharing platforms. Such results would, at the same time, increase credibility and trust in video-sharing platforms as an asset for digital mental health.
\section{Challenges}
Below are some of the challenges we envision when studying video-sharing platforms. We seek to foster discussion on how to tackle these challenges during our participation in the workshop.
\begin{itemize}
    \item Most video-sharing platforms, such as TikTok and YouTube, either have no official APIs or official APIs that are severely limiting in accessibility and capabilities. However, studying them through unofficial APIs and other technical solutions may bring about ethical as well as legal challenges.
    \item The multimodal nature of VSP content (music, speech, visuals, textual comments) makes it significantly more difficult to automatically detect misinformation compared to text-only content that is predominant on other social media platforms.  
    \item Assessing what is ground truth for mental health misinformation is challenging, as the scientific knowledge and landscape itself have differing perspectives at times, especially given the subjective experience of mental illness and the subjective nature of the psychiatric treatment. 
\end{itemize}

\section{Workshop Participation}
As a group, we have over a decade's worth of experience working toward understanding and fostering human well-being and safety on social media platforms. Participating in this workshop, we hope to contribute our insights on how to foster credibility, trust, and safety on video-sharing platforms. We also hope to gain insights from workshop participants on how best to solve the open questions we posed above, with a view toward design and technical solutions to mitigate mental health misinformation on video-sharing platforms.

\bibliographystyle{ACM-Reference-Format}
\bibliography{sample-base}


\begin{thebibliography}{27}


\ifx \showCODEN    \undefined \def \showCODEN     #1{\unskip}     \fi
\ifx \showDOI      \undefined \def \showDOI       #1{#1}\fi
\ifx \showISBNx    \undefined \def \showISBNx     #1{\unskip}     \fi
\ifx \showISBNxiii \undefined \def \showISBNxiii  #1{\unskip}     \fi
\ifx \showISSN     \undefined \def \showISSN      #1{\unskip}     \fi
\ifx \showLCCN     \undefined \def \showLCCN      #1{\unskip}     \fi
\ifx \shownote     \undefined \def \shownote      #1{#1}          \fi
\ifx \showarticletitle \undefined \def \showarticletitle #1{#1}   \fi
\ifx \showURL      \undefined \def \showURL       {\relax}        \fi
\providecommand\bibfield[2]{#2}
\providecommand\bibinfo[2]{#2}
\providecommand\natexlab[1]{#1}
\providecommand\showeprint[2][]{arXiv:#2}

\bibitem[Andrade et~al\mbox{.}(2014)]%
        {andrade2014barriers}
\bibfield{author}{\bibinfo{person}{Laura~Helena Andrade}, \bibinfo{person}{J
  Alonso}, \bibinfo{person}{Z Mneimneh}, \bibinfo{person}{JE Wells},
  \bibinfo{person}{A Al-Hamzawi}, \bibinfo{person}{G Borges},
  \bibinfo{person}{E Bromet}, \bibinfo{person}{Ronny Bruffaerts},
  \bibinfo{person}{G De~Girolamo}, \bibinfo{person}{R De~Graaf},
  {et~al\mbox{.}}} \bibinfo{year}{2014}\natexlab{}.
\newblock \showarticletitle{Barriers to mental health treatment: results from
  the WHO World Mental Health surveys}.
\newblock \bibinfo{journal}{\emph{Psychological medicine}}
  \bibinfo{volume}{44}, \bibinfo{number}{6} (\bibinfo{year}{2014}),
  \bibinfo{pages}{1303--1317}.
\newblock


\bibitem[Avella(2023)]%
        {avella2023tiktok}
\bibfield{author}{\bibinfo{person}{Holly Avella}.}
  \bibinfo{year}{2023}\natexlab{}.
\newblock \showarticletitle{“TikTok$\ne$ therapy”: Mediating mental health
  and algorithmic mood disorders}.
\newblock \bibinfo{journal}{\emph{New Media \& Society}}
  (\bibinfo{year}{2023}), \bibinfo{pages}{14614448221147284}.
\newblock


\bibitem[Basch et~al\mbox{.}(2022)]%
        {basch2022deconstructing}
\bibfield{author}{\bibinfo{person}{Corey~H Basch}, \bibinfo{person}{Lorie
  Donelle}, \bibinfo{person}{Joseph Fera}, {and} \bibinfo{person}{Christie
  Jaime}.} \bibinfo{year}{2022}\natexlab{}.
\newblock \showarticletitle{Deconstructing TikTok videos on mental health:
  cross-sectional, descriptive content analysis}.
\newblock \bibinfo{journal}{\emph{JMIR formative research}}
  \bibinfo{volume}{6}, \bibinfo{number}{5} (\bibinfo{year}{2022}),
  \bibinfo{pages}{e38340}.
\newblock


\bibitem[Devendorf et~al\mbox{.}(2020)]%
        {devendorf2020depression}
\bibfield{author}{\bibinfo{person}{Andrew Devendorf}, \bibinfo{person}{Ansley
  Bender}, {and} \bibinfo{person}{Jonathan Rottenberg}.}
  \bibinfo{year}{2020}\natexlab{}.
\newblock \showarticletitle{Depression presentations, stigma, and mental health
  literacy: A critical review and YouTube content analysis}.
\newblock \bibinfo{journal}{\emph{Clinical Psychology Review}}
  \bibinfo{volume}{78} (\bibinfo{year}{2020}), \bibinfo{pages}{101843}.
\newblock


\bibitem[Donzelli et~al\mbox{.}(2018)]%
        {donzelli2018misinformation}
\bibfield{author}{\bibinfo{person}{Gabriele Donzelli}, \bibinfo{person}{Giacomo
  Palomba}, \bibinfo{person}{Ileana Federigi}, \bibinfo{person}{Francesco
  Aquino}, \bibinfo{person}{Lorenzo Cioni}, \bibinfo{person}{Marco Verani},
  \bibinfo{person}{Annalaura Carducci}, {and} \bibinfo{person}{Pierluigi
  Lopalco}.} \bibinfo{year}{2018}\natexlab{}.
\newblock \showarticletitle{Misinformation on vaccination: A quantitative
  analysis of YouTube videos}.
\newblock \bibinfo{journal}{\emph{Human vaccines \& immunotherapeutics}}
  \bibinfo{volume}{14}, \bibinfo{number}{7} (\bibinfo{year}{2018}),
  \bibinfo{pages}{1654--1659}.
\newblock


\bibitem[Geeng et~al\mbox{.}(2020)]%
        {geeng2020fake}
\bibfield{author}{\bibinfo{person}{Christine Geeng}, \bibinfo{person}{Savanna
  Yee}, {and} \bibinfo{person}{Franziska Roesner}.}
  \bibinfo{year}{2020}\natexlab{}.
\newblock \showarticletitle{Fake news on Facebook and Twitter: Investigating
  how people (don't) investigate}. In \bibinfo{booktitle}{\emph{Proceedings of
  the 2020 CHI conference on human factors in computing systems}}.
  \bibinfo{pages}{1--14}.
\newblock


\bibitem[Grinberg et~al\mbox{.}(2019)]%
        {grinberg2019fake}
\bibfield{author}{\bibinfo{person}{Nir Grinberg}, \bibinfo{person}{Kenneth
  Joseph}, \bibinfo{person}{Lisa Friedland}, \bibinfo{person}{Briony
  Swire-Thompson}, {and} \bibinfo{person}{David Lazer}.}
  \bibinfo{year}{2019}\natexlab{}.
\newblock \showarticletitle{Fake news on Twitter during the 2016 US
  presidential election}.
\newblock \bibinfo{journal}{\emph{Science}} \bibinfo{volume}{363},
  \bibinfo{number}{6425} (\bibinfo{year}{2019}), \bibinfo{pages}{374--378}.
\newblock


\bibitem[Hunter(2022a)]%
        {hunter2022tiktokA}
\bibfield{author}{\bibinfo{person}{Tatum Hunter}.}
  \bibinfo{year}{2022}\natexlab{a}.
\newblock \bibinfo{title}{How to vet mental health advice on TikTok and
  Instagram}.
\newblock
\newblock
\urldef\tempurl%
\url{https://www.washingtonpost.com/technology/2022/10/03/tiktok-instagram-mental-health/}
\showURL{%
\tempurl}


\bibitem[Hunter(2022b)]%
        {hunter2022tiktokB}
\bibfield{author}{\bibinfo{person}{Tatum Hunter}.}
  \bibinfo{year}{2022}\natexlab{b}.
\newblock \bibinfo{title}{Online creators are de facto therapists for millions.
  it's complicated.}
\newblock
\newblock
\urldef\tempurl%
\url{https://www.washingtonpost.com/technology/2022/08/29/mental-health-tiktok-instagram/}
\showURL{%
\tempurl}


\bibitem[Hussein et~al\mbox{.}(2020)]%
        {hussein2020measuring}
\bibfield{author}{\bibinfo{person}{Eslam Hussein}, \bibinfo{person}{Prerna
  Juneja}, {and} \bibinfo{person}{Tanushree Mitra}.}
  \bibinfo{year}{2020}\natexlab{}.
\newblock \showarticletitle{Measuring misinformation in video search platforms:
  An audit study on YouTube}.
\newblock \bibinfo{journal}{\emph{Proceedings of the ACM on Human-Computer
  Interaction}} \bibinfo{volume}{4}, \bibinfo{number}{CSCW1}
  (\bibinfo{year}{2020}), \bibinfo{pages}{1--27}.
\newblock


\bibitem[Li et~al\mbox{.}(2020)]%
        {li2020youtube}
\bibfield{author}{\bibinfo{person}{Heidi Oi-Yee Li}, \bibinfo{person}{Adrian
  Bailey}, \bibinfo{person}{David Huynh}, {and} \bibinfo{person}{James Chan}.}
  \bibinfo{year}{2020}\natexlab{}.
\newblock \showarticletitle{YouTube as a source of information on COVID-19: a
  pandemic of misinformation?}
\newblock \bibinfo{journal}{\emph{BMJ global health}} \bibinfo{volume}{5},
  \bibinfo{number}{5} (\bibinfo{year}{2020}), \bibinfo{pages}{e002604}.
\newblock


\bibitem[Mena et~al\mbox{.}(2020)]%
        {mena2020misinformation}
\bibfield{author}{\bibinfo{person}{Paul Mena}, \bibinfo{person}{Danielle
  Barbe}, {and} \bibinfo{person}{Sylvia Chan-Olmsted}.}
  \bibinfo{year}{2020}\natexlab{}.
\newblock \showarticletitle{Misinformation on Instagram: The impact of trusted
  endorsements on message credibility}.
\newblock \bibinfo{journal}{\emph{Social Media+ Society}} \bibinfo{volume}{6},
  \bibinfo{number}{2} (\bibinfo{year}{2020}),
  \bibinfo{pages}{2056305120935102}.
\newblock


\bibitem[Pierri et~al\mbox{.}(2022)]%
        {pierri2022propaganda}
\bibfield{author}{\bibinfo{person}{Francesco Pierri}, \bibinfo{person}{Luca
  Luceri}, \bibinfo{person}{Nikhil Jindal}, {and} \bibinfo{person}{Emilio
  Ferrara}.} \bibinfo{year}{2022}\natexlab{}.
\newblock \showarticletitle{Propaganda and Misinformation on Facebook and
  Twitter during the Russian Invasion of Ukraine}.
\newblock \bibinfo{journal}{\emph{arXiv preprint arXiv:2212.00419}}
  (\bibinfo{year}{2022}).
\newblock


\bibitem[Serrano et~al\mbox{.}(2020)]%
        {serrano2020nlp}
\bibfield{author}{\bibinfo{person}{Juan Carlos~Medina Serrano},
  \bibinfo{person}{Orestis Papakyriakopoulos}, {and} \bibinfo{person}{Simon
  Hegelich}.} \bibinfo{year}{2020}\natexlab{}.
\newblock \showarticletitle{NLP-based feature extraction for the detection of
  COVID-19 misinformation videos on YouTube}. In
  \bibinfo{booktitle}{\emph{Proceedings of the 1st Workshop on NLP for COVID-19
  at ACL 2020}}.
\newblock


\bibitem[Shang et~al\mbox{.}(2021)]%
        {shang2021multimodal}
\bibfield{author}{\bibinfo{person}{Lanyu Shang}, \bibinfo{person}{Ziyi Kou},
  \bibinfo{person}{Yang Zhang}, {and} \bibinfo{person}{Dong Wang}.}
  \bibinfo{year}{2021}\natexlab{}.
\newblock \showarticletitle{A multimodal misinformation detector for covid-19
  short videos on tiktok}. In \bibinfo{booktitle}{\emph{2021 IEEE International
  Conference on Big Data (Big Data)}}. IEEE, \bibinfo{pages}{899--908}.
\newblock


\bibitem[Suarez-Lledo and Alvarez-Galvez(2021)]%
        {suarez2021prevalence}
\bibfield{author}{\bibinfo{person}{Victor Suarez-Lledo} {and}
  \bibinfo{person}{Javier Alvarez-Galvez}.} \bibinfo{year}{2021}\natexlab{}.
\newblock \showarticletitle{Prevalence of health misinformation on social
  media: systematic review}.
\newblock \bibinfo{journal}{\emph{Journal of medical Internet research}}
  \bibinfo{volume}{23}, \bibinfo{number}{1} (\bibinfo{year}{2021}),
  \bibinfo{pages}{e17187}.
\newblock


\bibitem[Swire-Thompson and Lazer(2019)]%
        {swire2019public}
\bibfield{author}{\bibinfo{person}{Briony Swire-Thompson} {and}
  \bibinfo{person}{David Lazer}.} \bibinfo{year}{2019}\natexlab{}.
\newblock \showarticletitle{Public health and online misinformation: challenges
  and recommendations.}
\newblock \bibinfo{journal}{\emph{Annual review of public health}}
  \bibinfo{volume}{41} (\bibinfo{year}{2019}), \bibinfo{pages}{433--451}.
\newblock


\bibitem[Tal and Torous(2017)]%
        {tal2017digital}
\bibfield{author}{\bibinfo{person}{Amir Tal} {and} \bibinfo{person}{John
  Torous}.} \bibinfo{year}{2017}\natexlab{}.
\newblock \showarticletitle{The digital mental health revolution: Opportunities
  and risks.}
\newblock  (\bibinfo{year}{2017}).
\newblock


\bibitem[Team(2022)]%
        {plushcare2022tiktok}
\bibfield{author}{\bibinfo{person}{PlushCare~Content Team}.}
  \bibinfo{year}{2022}\natexlab{}.
\newblock \bibinfo{title}{How accurate is mental health advice on TikTok?}
\newblock
\newblock
\urldef\tempurl%
\url{https://plushcare.com/blog/tiktok-mental-health/}
\showURL{%
\tempurl}


\bibitem[Tomlein et~al\mbox{.}(2021)]%
        {tomlein2021audit}
\bibfield{author}{\bibinfo{person}{Matus Tomlein}, \bibinfo{person}{Branislav
  Pecher}, \bibinfo{person}{Jakub Simko}, \bibinfo{person}{Ivan Srba},
  \bibinfo{person}{Robert Moro}, \bibinfo{person}{Elena Stefancova},
  \bibinfo{person}{Michal Kompan}, \bibinfo{person}{Andrea Hrckova},
  \bibinfo{person}{Juraj Podrouzek}, {and} \bibinfo{person}{Maria Bielikova}.}
  \bibinfo{year}{2021}\natexlab{}.
\newblock \showarticletitle{An audit of misinformation filter bubbles on
  YouTube: Bubble bursting and recent behavior changes}. In
  \bibinfo{booktitle}{\emph{Proceedings of the 15th ACM Conference on
  Recommender Systems}}. \bibinfo{pages}{1--11}.
\newblock


\bibitem[Torous et~al\mbox{.}(2020)]%
        {torous2020digital}
\bibfield{author}{\bibinfo{person}{John Torous}, \bibinfo{person}{Keris~J{\"a}n
  Myrick}, \bibinfo{person}{Natali Rauseo-Ricupero}, \bibinfo{person}{Joseph
  Firth}, {et~al\mbox{.}}} \bibinfo{year}{2020}\natexlab{}.
\newblock \showarticletitle{Digital mental health and COVID-19: using
  technology today to accelerate the curve on access and quality tomorrow}.
\newblock \bibinfo{journal}{\emph{JMIR mental health}} \bibinfo{volume}{7},
  \bibinfo{number}{3} (\bibinfo{year}{2020}), \bibinfo{pages}{e18848}.
\newblock


\bibitem[Wainberg et~al\mbox{.}(2017)]%
        {wainberg2017challenges}
\bibfield{author}{\bibinfo{person}{Milton~L Wainberg}, \bibinfo{person}{Pamela
  Scorza}, \bibinfo{person}{James~M Shultz}, \bibinfo{person}{Liat Helpman},
  \bibinfo{person}{Jennifer~J Mootz}, \bibinfo{person}{Karen~A Johnson},
  \bibinfo{person}{Yuval Neria}, \bibinfo{person}{Jean-Marie~E Bradford},
  \bibinfo{person}{Maria~A Oquendo}, {and} \bibinfo{person}{Melissa~R
  Arbuckle}.} \bibinfo{year}{2017}\natexlab{}.
\newblock \showarticletitle{Challenges and opportunities in global mental
  health: a research-to-practice perspective}.
\newblock \bibinfo{journal}{\emph{Current psychiatry reports}}
  \bibinfo{volume}{19} (\bibinfo{year}{2017}), \bibinfo{pages}{1--10}.
\newblock


\bibitem[Wang et~al\mbox{.}(2005)]%
        {wang2005failure}
\bibfield{author}{\bibinfo{person}{Philip~S Wang}, \bibinfo{person}{Patricia
  Berglund}, \bibinfo{person}{Mark Olfson}, \bibinfo{person}{Harold~A Pincus},
  \bibinfo{person}{Kenneth~B Wells}, {and} \bibinfo{person}{Ronald~C Kessler}.}
  \bibinfo{year}{2005}\natexlab{}.
\newblock \showarticletitle{Failure and delay in initial treatment contact
  after first onset of mental disorders in the National Comorbidity Survey
  Replication}.
\newblock \bibinfo{journal}{\emph{Archives of general psychiatry}}
  \bibinfo{volume}{62}, \bibinfo{number}{6} (\bibinfo{year}{2005}),
  \bibinfo{pages}{603--613}.
\newblock


\bibitem[Xia et~al\mbox{.}(2022)]%
        {xia2022millions}
\bibfield{author}{\bibinfo{person}{Haijun Xia}, \bibinfo{person}{Hui~Xin Ng},
  \bibinfo{person}{Zhutian Chen}, {and} \bibinfo{person}{James Hollan}.}
  \bibinfo{year}{2022}\natexlab{}.
\newblock \showarticletitle{Millions and Billions of Views: Understanding
  Popular Science and Knowledge Communication on Video-Sharing Platforms}. In
  \bibinfo{booktitle}{\emph{Proceedings of the Ninth ACM Conference on
  Learning@ Scale}}. \bibinfo{pages}{163--174}.
\newblock


\bibitem[Xie et~al\mbox{.}(2022)]%
        {xie2022interpretable}
\bibfield{author}{\bibinfo{person}{Jiaheng Xie}, \bibinfo{person}{Yidong Chai},
  {and} \bibinfo{person}{Xiao Liu}.} \bibinfo{year}{2022}\natexlab{}.
\newblock \showarticletitle{An Interpretable Deep Learning Approach to
  Understand Health Misinformation Transmission on YouTube}. In
  \bibinfo{booktitle}{\emph{Proceedings Of The 55th Hawaii International
  Conference On System Sciences}}.
\newblock


\bibitem[Yeung et~al\mbox{.}(2022)]%
        {yeung2022tiktok}
\bibfield{author}{\bibinfo{person}{Anthony Yeung}, \bibinfo{person}{Enoch Ng},
  {and} \bibinfo{person}{Elia Abi-Jaoude}.} \bibinfo{year}{2022}\natexlab{}.
\newblock \showarticletitle{TikTok and attention-deficit/hyperactivity
  disorder: a cross-sectional study of social media content quality}.
\newblock \bibinfo{journal}{\emph{The Canadian Journal of Psychiatry}}
  \bibinfo{volume}{67}, \bibinfo{number}{12} (\bibinfo{year}{2022}),
  \bibinfo{pages}{899--906}.
\newblock


\bibitem[Zhu et~al\mbox{.}(2020)]%
        {zhu2020health}
\bibfield{author}{\bibinfo{person}{Chengyan Zhu}, \bibinfo{person}{Xiaolin Xu},
  \bibinfo{person}{Wei Zhang}, \bibinfo{person}{Jianmin Chen}, {and}
  \bibinfo{person}{Richard Evans}.} \bibinfo{year}{2020}\natexlab{}.
\newblock \showarticletitle{How health communication via Tik Tok makes a
  difference: A content analysis of Tik Tok accounts run by Chinese provincial
  health committees}.
\newblock \bibinfo{journal}{\emph{International journal of environmental
  research and public health}} \bibinfo{volume}{17}, \bibinfo{number}{1}
  (\bibinfo{year}{2020}), \bibinfo{pages}{192}.
\newblock


\end{thebibliography}

\end{document}